\documentclass[prl,twocolumn,letterpaper,superscriptaddress]{revtex4-1}
\usepackage{bm,graphicx,graphics,amsmath,amssymb,bm,epsfig,color,mathrsfs}
\usepackage{euscript,tabularx}
\usepackage{longtable}
\usepackage{tipa}
\usepackage{frcursive}

\def\<{\left\langle}
\def\>{\right\rangle}
\def\unit#1{{\hat{\bm#1}}}
\def\op#1{{\widehat{\bm#1}}}

\newcommand{\mathfrc}[1]{\text{\raisebox{0.3ex}{\footnotesize \cursive#1}}}
\def\zi{\mathfrc{z}}
\def\wi{\mathit{w}}

\begin{document}

\title{Probing the anharmonicity of the potential well for magnetic
  vortex core in a nanodot}

\author{O.V. Sukhostavets} \affiliation{Departamento de F{\'i}sica
  de Materiales, Universidad del Pais Vasco, 20018 San Sebastian,
  Spain}

\author{B. Pigeau}
\affiliation{Service de Physique de l'\'Etat Condens\'e (CNRS URA
  2464), CEA Saclay, 91191 Gif-sur-Yvette, France}

\author{S. Sangiao} \affiliation{Service de Physique de l'\'Etat
  Condens\'e (CNRS URA 2464), CEA Saclay, 91191 Gif-sur-Yvette,
  France}

\author{G. de Loubens} \affiliation{Service de Physique de l'\'Etat
  Condens\'e (CNRS URA 2464), CEA Saclay, 91191 Gif-sur-Yvette,
  France}

\author{V.V. Naletov} \affiliation{Service de Physique de l'\'Etat
  Condens\'e (CNRS URA 2464), CEA Saclay, 91191 Gif-sur-Yvette,
  France} \affiliation{Institute of Physics, Kazan Federal University,
  Kazan 420008, Russian Federation}

\author{O. Klein} \email[Corresponding author:]{ oklein@cea.fr}
\affiliation{Service de Physique de l'\'Etat Condens\'e (CNRS URA
  2464), CEA Saclay, 91191 Gif-sur-Yvette, France}

\author{K. Mitsuzuka}
\author{S. Andrieu}
\author{F. Montaigne}
\affiliation{Institut Jean Lamour, UMR CNRS 7198, Universit{\'e} de Lorraine, 54
506 Nancy, France}

\author{K.Y. Guslienko} \affiliation{Departamento de F{\'i}sica
  de Materiales, Universidad del Pais Vasco, 20018 San Sebastian,
  Spain} \affiliation{IKERBASQUE, The Basque Foundation for Science,
  48011 Bilbao, Spain}

\date{\today}

\begin{abstract}
  The anharmonicity of the potential well confining a magnetic vortex
  core in a nanodot is measured dynamically with a Magnetic Resonance
  Force Microscope (MRFM). The stray field of the MRFM tip is used to
  displace the equilibrium core position away from the nanodot
  center. The anharmonicity is then inferred from the relative
  frequency shift induced on the eigen-frequency of the vortex core
  translational mode. An analytical framework is proposed to extract
  the anharmonic coefficient from this variational approach. Traces of
  these shifts are recorded while scanning the tip above an isolated
  nanodot, patterned out of a single crystal FeV film.  We observe
  $+10$\% increase of the eigen-frequency when the equilibrium
  position of the vortex core is displaced to about one third of its
  radius. This calibrates the tunability of the gyrotropic mode by
  external magnetic fields.
\end{abstract}

\maketitle

There has been recently a renewed interest, both theoretical and
experimental, in the problem of nonlinear (NL) magnetization dynamics
inside confined nanostructures \cite{Slavin2009}. NL phenomena are
responsible for the creation of novel dynamical objects
\cite{Mohseni2013}, analogs of dynamical solitons. They also set the
figure of merit of spintronics devices, \textit{e.g.} the spectral
purity and tuning sensitivity of spin transfer torque nano-oscillators
\cite{Slavin2009}. On the theoretical side, predictions on the
amplitude of the NL coefficients have been found to be extremely
difficult to compute beyond the uniformly magnetized ground state. The
difficulty raises both from the magneto-dipolar field, which
introduces a non-local interaction, and from the kinetic part of the
effective field (or gauge field), which modifies the texture of the
magnetic configuration. On the experimental side, the most promising
findings have been discovered on non-uniform ground states, such as
magnetic vortex existing in ferromagnetic nanodot. Vortices have
stimulated the emergence of higher performance microwave oscillators
using isolated \cite{Pribiag2007} or dipolarly coupled
\cite{Locatelli2011} nanodot, or for future magnetic memories by
allowing the resonant switching of the magnetic configuration
\cite{Pigeau2010}.

Magnetic vortex corresponds to a curling in-plane magnetization
spatial distribution leaving a nanometric in size core region ($\sim$
the exchange length), where the magnetization is pointing
out-of-plane. The lowest energy mode is a translational (or
gyrotropic) mode of the vortex core position $\bm X$, expressed here
in reduced unit of the dot radius. Its properties are governed by the
magnetostatic potential well $W^\text{(M)}$ in which the core
evolves. For a circular nanodot, the magnetostatic energy is isotropic
in the dot plane and it can be written as a series expansion of even
powers of the dimensionless $\bm X$ \cite{Dussaux2012}:
\begin{equation} \label{eq:potential} W^\text{(M)} = W^\text{(M)}_0 +
  \frac 1 2 \kappa |{\bm X}|^2 + \frac 1 4 \kappa^\prime |{\bm X}|^4
  +{\cal{O}}( |{\bm X}|^6)\,,
\end{equation}
At the present, only the parabolicity of the confinement, $\kappa$,
has been well characterized experimentally and the measured value is
in agreement with theoretical predictions \cite{Novosad2005}.
This is not the case for the higher order terms and there is no
consensus yet on the order of magnitude or the sign of the anharmonic
coefficient $\lambda \equiv \kappa'/\kappa$ afferent to the
depolarisation effect of a displaced vortex inside a large planar
circular nanodot. The asymptotic limit of large radius is the relevant
aspect ratio to test the dipole dominating limit and the circular
symmetry is necessary to avoid the additional complexity of
non-isotropic potential found for example in square shaped elements
\cite{Drews2012}. Up to now, attempts to measure $\lambda$ in circular
nanodot using large rf excitation have so far lead to inconsistent
results between experiments \cite{Pigeau2011} (red shift) and theory
\cite{Gaididei2010} (blue shift). The measurement of $\lambda$ through
a variational approach, consisting in studying the small change of
oscillation period when a large static displacement of the vortex core
equilibrium position is produced, has so far failed too: this
counterpart of large rf oscillation has mostly revealed the potential
well inhomogeneities leading to pinning of the core
\cite{Chen2012,Burgess2013}.

In this work, we report on an experimental measurement of $\lambda$ in
a large planar circular nanodot using a Magnetic Resonance Force
Microscope (MRFM).  All the experimental measurements are performed on
an individual nanodisk of thickness $t=26.7$~nm thick and nominal
radius $R=300$~nm, patterned out of a single crystal FeV film. Only
the perfect crystalline structure ensures an unpinned displacement of
the vortex core throughout the sample volume. We rely here on the
non-uniform stray field of the magnetic tip of the MRFM to displace
the vortex core away from the nanodot center. The anharmonic
coefficient is then inferred from the measurement of the relative
variation of the eigen-frequency of the gyrotropic mode as a function
of the tip displacement.

\begin{figure}
  \includegraphics[width=8.5cm]{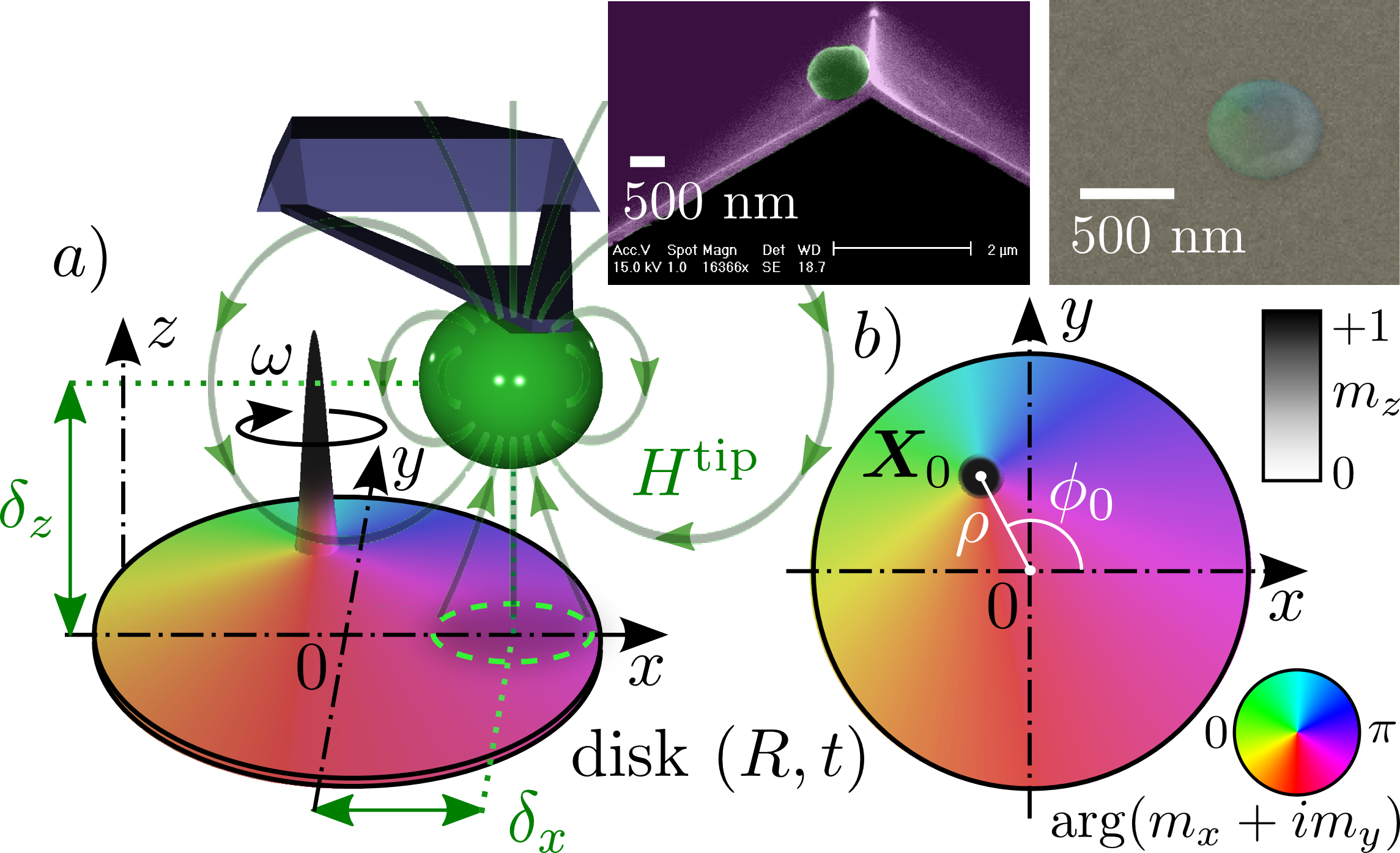}
  \caption{(Color online) a) Side and b) top views of the experimental
    setup: the stray field of an MRFM tip is used to displace the
    static position $\bm{X}_0$ of the vortex core. The magnetic vortex
    state is shown in a bi-variate colormap of $\bm{m}=\bm{M}/M_s$
    (amplitude-phase $\leftrightarrow$ luminance-hue). The insets are
    microscopy images of the magnetic tip and disk sample.}
  \label{FIG1}
\end{figure}

We start first with a description of the experimental setup
\cite{Klein2008} (FIG.\ref{FIG1}). The right inset shows an image of
the sample: a circular nanodot, which is patterned by standard
lithography and ion-milling techniques from an extended film of Fe-V
(10\% V)\cite{Mitsuzuka2012} with magnetization $4 \pi M_s = 17$~kG. A
magnetic tip is brought in the vicinity of the sample (left image).
The tip consists of a soft Fe particle glued at the apex of
micro-cantilever.  The MRFM is placed inside a superconducting coil
magnet, which produces an homogenous bias magnetic field $\bm {H}_0$
of 6~kOe along the $z$-direction (parallel to the normal of the
disk). The value of ${H}_0$ is chosen to be strong enough to magnetize
the MRFM tip close to its saturation value, while remaining weak
enough compared to the saturation field of the nanodot to preserve the
vortex ground state inside the sample.  At ${H}_0$ the tip stray field
is $\bm{H}^\text{tip} (\bm {r}) = -\nabla (\bm{\mu}_\text{tip}\cdot
\bm{r} / r^3)$, the dipolar field generated by a point magnetic moment
${\mu}_\text{tip} = 4\times 10^{-10}$~emu oriented along $\unit z$
\cite{Pigeau2012}. Effects of perpendicular magnetic field on magnetic
vortex are well established \cite{Ivanov2002,Loubens2009}: the
in-plane spins are tilted towards the applied field producing hereby a
decrease \footnote{At $H_0=6$~kOe, the spins are tilted out-of-plane
  by about 20$^\circ$ generating a 7\% decrease of the in-plane
  component of the magnetization outside the vortex core.} of the
in-plane component of the magnetization outside the vortex core (cone
state \cite{Ivanov2002}).

\begin{figure}
  \includegraphics[width=8.5cm]{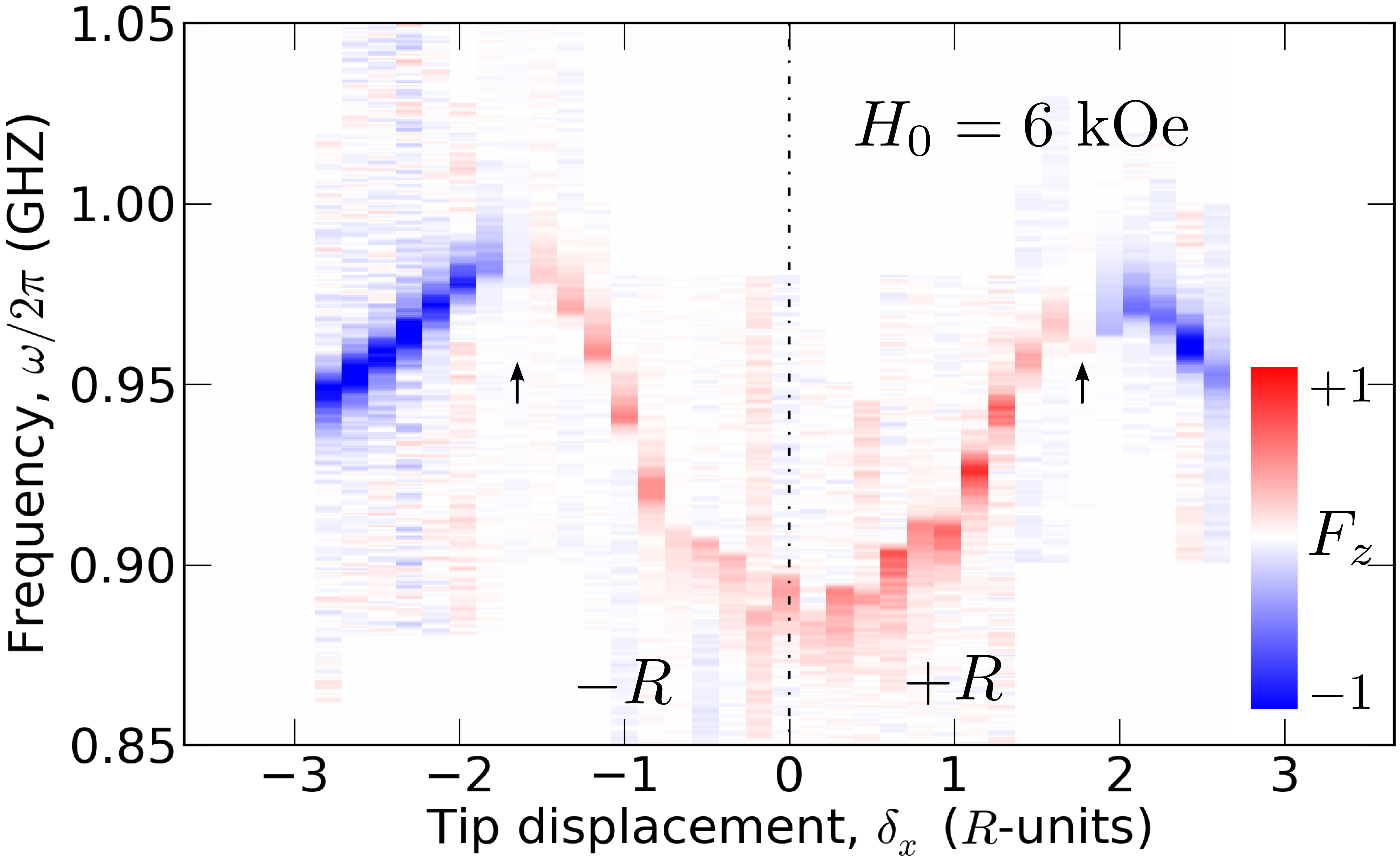}
  \caption{(Color online) Density plot showing the experimentally
    measured variation of the eigen-frequency of the gyrotropic mode
    upon a lateral displacement, $\delta_x$, of the MRFM tip at fixed
    $\delta_y=0$ and nominal $\delta_z=3.0$. A red-blue colormap shows
    the sign of the force acting on the cantilever. The arrows
    indicate where the change of polarity occurs. }
  \label{FIG2}
\end{figure}

We then proceed to the measurement of the variation of the excitation
spectrum of the gyrotropic mode when the tip is scanned by $\pm
0.85\mu$m along the $x$-direction by steps of 50~nm (see
FIG.\ref{FIG2}). The scan height is $0.9 \pm 0.05 \mu$m
\footnote{Although the use of piezo-actuators allows ultra-precise
  displacement of the micro-cantilever, the value of $\delta_z$ has
  inherently some uncertainty as it corresponds to the free axis of
  the cantilever.}  above the nanodot or $\delta_z=3.0 \pm 0.15$ in
reduced units of $R$ (hereafter all spatial displacements are
expressed in units of the dot radius $R$.)  Placing the tip at the
origin ($\delta_x=0$ or on the axis of the disk), attracts the vortex
core at the center of the nanodot. From there, lateral displacement of
the magnetic tip produces a vector shift ${\bm X}_0$ of the vortex
core equilibrium position from the dot center. The process is driven
by the growth of the in-plane domain parallel to the in-plane
component of the tip stray field. We observe in FIG.\ref{FIG2} that
the eigen-frequency \emph{increases} (blue shift) upon increasing
$\left | {\bm X}_0 \right |$. Noting that the frequency shift is
symmetric and isotropic (the signature that the intrinsic potential is
being probed) we find that the vortex core dynamics can be tuned on a
relative large range ($\sim$ 10\%), hereby demonstrating that the
magnetostatic potential must be anharmonic, since a purely parabolic
shape would have lead to a frequency independent behavior on ${\bm
  X}_0$.

Analysis of the amplitude of the MRFM signal gives a hint on the
amount of displacement $ \left|\bm{X}_0 \right|$ achieved during a
scan. The MRFM signal corresponds to the difference of vertical force
$\Delta F_z$ acting on the cantilever when the vortex motion is
excited. Defining $\bm{m} = \bm{M}/M_s $ the reduced magnetization
vector, the gyromotion produces a diminution of the spontaneous
magnetization along the local equilibrium direction $ \Delta m_i =
\frac 1 2 \left| \partial_X m_i + j\, \partial_Y m_i \right
|^2_{\bm{X}=\bm{X}_0}$, that mostly occurs outside the core region
\cite{Guslienko2008a}. The generated force can be then calculated from
the reaction force $-\Delta F_z = V M_s \langle g_{zi} \Delta m_i
\rangle$ acting on the nanodot due to the gradient tensor of the tip:
$\op g = \nabla \bm{H}^\text{tip}$. Cartesian tensor notation is used
here, with repeated indices being assumed summed. The chevron bracket
indicates that the enclosed quantity is averaged over the volume $V$
of the nanodot. The red-blue colormap in FIG.\ref{FIG2} codes the
amplitude MRFM signal: red (blue) means attractive (repulsive) force.
Small arrows at $\delta_x \approx \pm 1.7$ indicate the compensation
point of the force: where the change of sign occurs. This distance is
about twice smaller than the one required to change the sign of the
force in the saturated state (contribution dominated by $\langle
g_{zz} \Delta m_z \rangle$). The difference is interpreted as due to
the translation of the core position transversally to the tip position
(along the $y$-axis). The growth of this domain generates a repulsive
vertical force on the tip trough the cross gradient term $\langle
g_{zx} \Delta m_x \rangle$. We shall thus use the position of the
compensation point to calibrate precisely the amplitude of the
displacement of the vortex core.

Our next step is to develop an analytical framework allowing the
extraction of $\lambda$ from this variational study. The first stage
of this analysis is to calculate the equilibrium position $\bm
{X}_0=(X_0,Y_0)$ (here $X_{0}$ and $Y_{0}$ are the two in-plane
cartesian coordinates) by minimizing the total energy
$W=W^\text{(M)}+W^\text{(H)}$, the sum of $W^\text{(M)}$, the
magnetostatic self-energy of the vortex ground state which confines
the vortex core to the center of the nanodot and $W^\text{(H)}= - V
M_s \langle \bm m \cdot \bm H\rangle$, the Zeeman energy which
represents the interaction with the external magnetic field
$\bm{H}=\bm{H}_0 + \bm{H}^\text{tip}$ and is responsible for the
displacement $\bm {X}_0$. The magnetic configuration inside the
nanodot $m_x + j\, m_y = 2 \wi/(1+ \wi \wi^\ast)$ is conveniently
described by a conformal mapping of the complex variable $\wi$ ($\ast$
indicating the complex conjugate), which is a piecewise function of
the complex position $\zi=(x+ j \, y)/R$ with $\wi=f(\zi)/|f(\zi)|$
outside the vortex core region and $\wi=f(\zi)$ inside. The function
$f$ captures the texture of the spatial configuration. To calculate
the new equilibrium position $\mathcal{Z}_0=(X_0+ j \, Y_0)$, it is
appropriate to describe the dot magnetization, $\bm m$, by the rigid
vortex model (RVM) \cite{Guslienko2001}, written as $f(\zi) = \pm j \,
(\zi-\mathcal{Z}_0)/r_c$.  Here $r_c=R_c/R$ denotes the core radius
$R_c$ in reduced unit of $R$ and the $\pm$ sign depends on the
chirality of the vortex. The static displacement $\bm {X}_0$ is then
obtained by minimizing the total energy \footnote{All energies can be
  approximately calculated by running the integrand solely on the
  nanodot volume outside the vortex core region.} with the analytical
expression of $W^{\text{(M)}}$ obtained by the RVM. In the RVM, the
magnetostatic energy is generated by the surface magnetic charges
$\sigma$ located at the circumference of the disk (the volume charges
$\nabla \cdot \bm{M}$ are absent). The confinement potential follows
from the integral $W^{\text{(M)}} = \frac{1}{2} \int d\phi \int d\phi'
\sigma(\phi) \sigma(\phi') / \sqrt{2(1-\cos(\phi - \phi'))}$ where the
integration is taken over the disk periphery and $\sigma$ is given by:
\begin{equation} \label{eq:sigma}
\sigma(\phi) = + M_s
  \frac{-|\bm {X}_0| \sin(\phi - \phi_{0})}{\sqrt{1- 2
      |\bm {X}_0| \cos(\phi - \phi_{0})
      +|\bm {X}_0|^2 }} \,,
\end{equation}
$\phi_0$ is the azimuthal direction of the vortex equilibrium position
measured from the averaged in-plane bias field direction (here
$x$-axis). 

\begin{figure}
  \includegraphics[width=8.5cm]{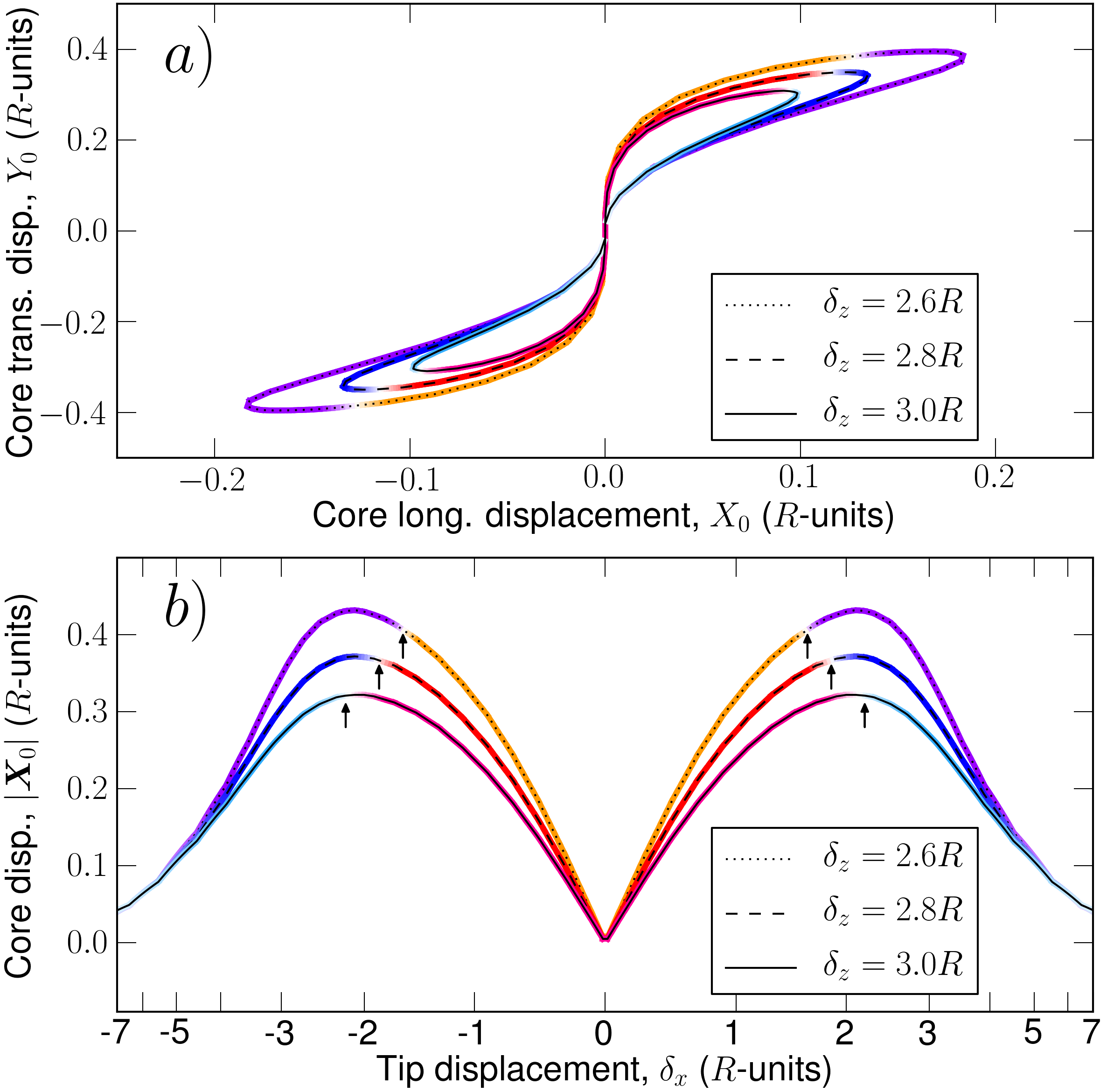}
  \caption{(Color online) a) Trajectory of the vortex core ${\bm X}_0=
    (X_0,Y_0)$ during an implicit lateral scan of the tip $\delta_x
    \in [-8,+8]$ at 3 different heights, $\delta_z$. b) Norm of the
    displacement vector, $ \left|\bm{X}_0 \right|$, as a function of
    the tip position, $\delta_x$.}
  \label{FIG3}
\end{figure}

The implicit trajectory of $\bm {X}_0$ is shown in FIG.\ref{FIG3}a for
three different heights $\delta_z$ around the nominal value. As
expected the in-plane components of the tip magnetic field displace
the vortex core mainly along the $y$-axis. The displacement along
$x$-axis is approximately twice smaller.  The resulting displacement
distance $|\bm{X}_0|$ as a function of $\delta_x$ is shown in
FIG.\ref{FIG3}b. We use here a skewed scale on the abcisse to show the
behavior when $\delta_x \gg 1$. We have also calculated the
corresponding dipolar force produced on the tip. The result is coded
in the colormap using the same convention as in FIG.\ref{FIG2}. We
have placed small arrows at the compensation points. Since decreasing
the scan height increases the amplitude of $|\bm{X}_0|$, we find that
the position of the arrows sensitively depends on $\delta_z$. Varying
$\delta_z$ in the experimental error bars $[2.6,3.0]$ displaces the
compensation point by $\pm 0.3 \cdot R$ (or $\pm 100$~nm) around the
mean value $\delta_x=1.8$, in agreement with the experimental data. We
shall use this marker to evaluate the uncertainty window of
$|\bm{X}_0|$ in our experiment.


The second stage of this analysis is to perform a linearization of the
vortex equation of motion to a cyclic excitation field. The
instantaneous response $\bm{X}=(X,Y)$ is decomposed into the static
component $\bm{X}_0$, calculated previously, and a dynamic component
$\bm{\xi}=\bm{X} -\bm{X}_0$ representing the small oscillating
deviation of the vortex core position from its equilibrium
\footnote{We estimate that $\xi=0.07$ in our experiment, where the
  microwave field strength is $h_\text{rf}=0.6$~Oe.}. In the dynamical
case, the dipolar pinning imposes a precession node at the dot
circumference \cite{Guslienko2008a}. It implies that the dynamical
magnetization comes from the variation, $\partial_X {\bm m} +
j\, \partial_Y {\bm m}$, of a magnetic configuration that has no
radial component at the dot border. Therefore, to calculate the
frequency of the small dynamic vortex displacement $\bm \xi$, it is
appropriate to use the surface charges free model or two vortex ansatz
(TVA) written as $f(\zi) = \mp j \, \frac{1}{r_c}
(\zi-\mathcal{Z})(\zi\mathcal{Z}^\ast-1)/(1+|\mathcal{Z}|^2)$
\cite{Guslienko2002} with $\mathcal{Z}=(X+ j \, Y)$. In our notation,
the dampingless Thiele equation simply writes $ \bm{G} \times
\dot{\bm{\xi}} = \partial W/\partial \bm{\xi} $, where $W$ is the
total energy and $ \left | \bm{G} \right|=2 \pi M_s t / \gamma$ is the
gyrovector \cite{Guslienko2008} (the dot is the short hand notation
for the time derivative and $\gamma$ is the gyromagnetic
ratio). Linearization around $\bm {X}_0$ yields the gyrotropic angular
frequency
\begin{equation} \label{eq:freq} 
  \omega^2 = \frac{ K_{xx} K_{yy}-K_{xy}^2}{G^2} \,,
  \ \text{with} \ 
  K_{ij} \equiv \left . \frac{\partial^2 W}{\partial {\xi}_i \partial
      {\xi}_j} \right |_{\bm{X}=\bm{X}_0}
\end{equation}
being the stiffness of the vortex core to small displacements in both
the $i$ and $j$ directions. Distinction between different cartesian
directions is necessary once the trajectory becomes elliptical.  This
is precisely, what occurs when $\bm{X}_0 \gg \bm \xi$: the amplitude
of the $\xi$-component along $\bm{X}_0$ differs from the amplitude of
the $\xi$-component perpendicular to $\bm{X}_0$ (short axis of the
ellips is along the radial direction). The degree of ellipticity is
determimed by the anharmonic contribution $\lambda \left | \bm{X}_0
\right |^2$.  This is in contrast to the opposite limit $\bm{X}_0 \ll
\bm \xi$, where the trajectory corresponds to a large amplitude
circular vortex core motion around the nanodot center
\cite{Dussaux2012}.

To calculate the different tensor elements of the stiffness $K_{ij} =
K^\text{(M)}_{ij} + K^\text{(H)}_{ij}$, one must decompose it in two
contributions corresponding respectively to the magnetostatic and
Zeeman energies. The first order value of the TVA magnetostatic
stiffness, $\kappa$, has been already expressed analytically
\cite{Guslienko2006}. The analytical expression of the anharmonic
correction is obtained by inserting Eq.(\ref{eq:potential}) in
Eq.(\ref{eq:freq}) and it leads to a simplified expression $
K^\text{(M)}_{ij} = \kappa \left . \left( \delta_{ij} + \lambda |\bm
    {X}|^2 \delta_{ij} + 2 \lambda X_{i} X_{j} \right ) \right
|_{\bm{X} = \bm{X}_0}$. It turns out that the Zeeman stiffness can be
neglected. Indeed, it can be shown that the tip stray field produces
no Zeeman stiffness along the diagonal elements
($K^\text{(H)}_{ii}=0$). Only the cross-terms $K^\text{(H)}_{xy}\neq0$
are non-vanishing but they represent a negligible correction ($<$
3\%). We thus find that at $\bm {X}_0 = 0$ and $H_z=0$,
Eq.(\ref{eq:freq}) simplifies to the well known expression
$\omega(0,0)={\kappa}/{G}$ \cite{Guslienko2002}.  At $\bm {X}_0 = 0$
and $H_z\neq 0$, the stiffness of the magnetostatic potential is
renormalized by the in-plane magnetization projection of the cone
state and one obtains ${\omega(0,H_z)}/\omega(0,0) = 1 + H_z/(4 \pi
M_s) $ \cite{Loubens2009}. In the general case $\bm {X}_0 \neq 0$ and
$H_z\neq 0$, the relative frequency shift reduces to the following
analytical expression:
\begin{equation} \label{eq:omegaG} \frac{\omega(\bm
    {X_0},H_z)}{\omega(0,H_z)} = 1 + 2 \lambda \left| \bm {X_0}
  \right|^2 + {\cal{O}}( |{\bm X}|^4)\,.
\end{equation}
Notice that the prefactor of 2 multiplying $\lambda$ is specific to
the limit $\bm{X}_0 \gg \bm \xi$.

\begin{figure}
  \includegraphics[width=8.5cm]{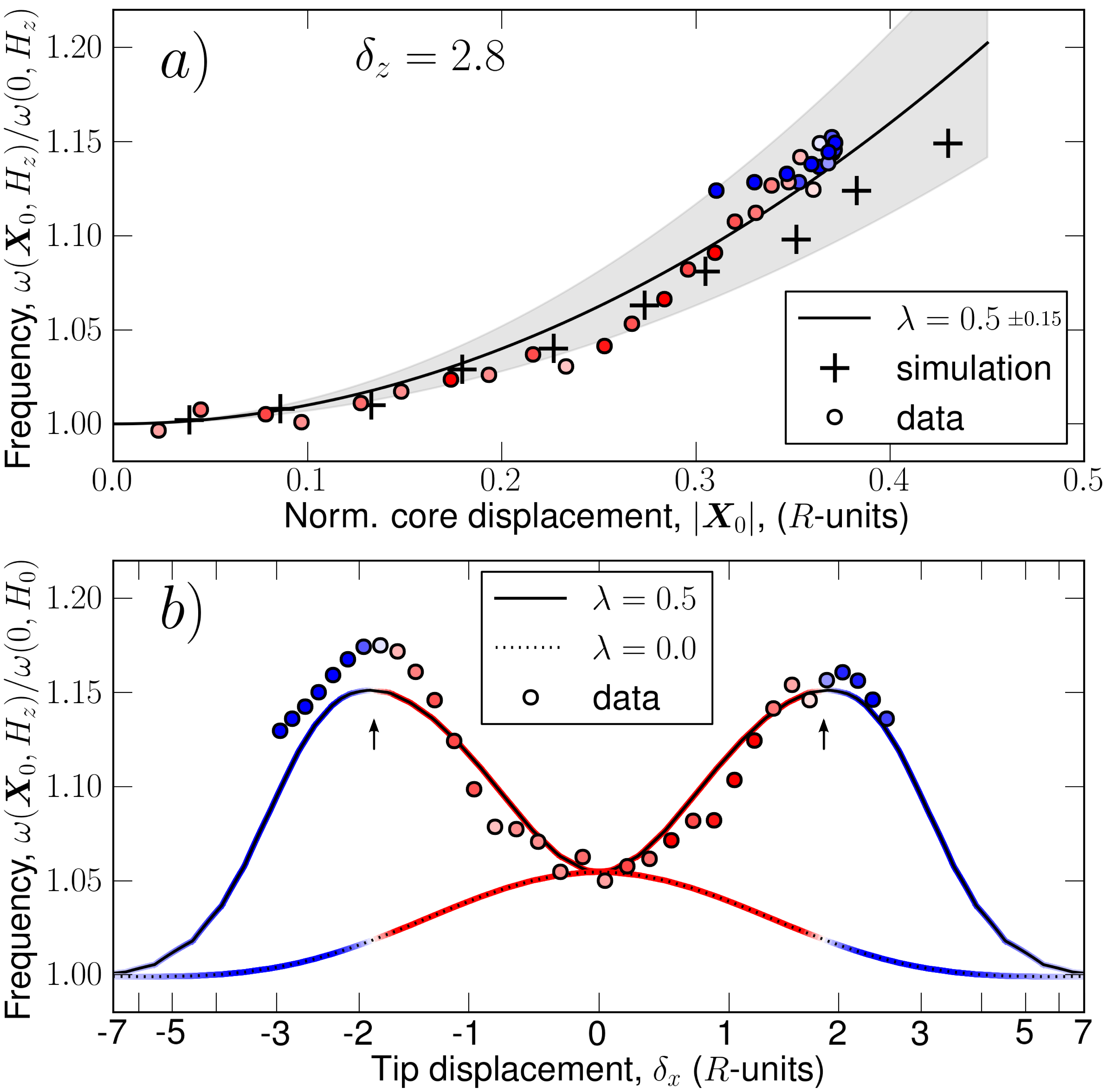}
  \caption{(Color online) a) Plot of the relative variation of
    eigen-frequency of the gyrotropic mode as a function of $|
    \bm{X}_0 |$. A fit by a parabola yields $\lambda=0.5\pm0.15$. The
    cross are the results of micromagnetic simulations b) Relative
    variation of the eigen-frequency as a function of $\delta_x$. The
    lines show the analytically predicted behavior for a vanishing (0)
    and finite (0.5) anharmonic coefficient. }
  \label{FIG4}
\end{figure}

The next step is to plot in FIG.\ref{FIG4}a the experimental data
extracted from FIG.\ref{FIG2}, renormalized by the predicted
dependence of ${\omega(0,H_z)}$, as a function of the calculated
$|\bm{X}_0|$ during a lateral scan of the tip at fixed
$\delta_z=2.8$. Fitting the data of FIG.\ref{FIG4}a with a parabola
(solid line) yields an average curvature $\lambda=0.5$.  We have
plotted in FIG.\ref{FIG4}b the experimentally measured relative
frequency normalized by $\omega(0,H_0)$. The latter quantity is
inferred experimentally by studying the decay of $\omega$ upon
increasing $\delta_z$, while keeping the tip on the symmetry axis
($\delta_x=\delta_y=0$): a fit of the decay behavior yields the
asymptotic value $\omega(0,H_0)$. In FIG.\ref{FIG4}b, the data point
are colored according to the colormap associated with the amplitude of
the force. For comparison, we have also plotted the predicted
variation of $\omega$ by Eq.(\ref{eq:omegaG}) as a function of
$\delta_x$ for two values of $\lambda$. Setting $\lambda=0$ in
Eq.(\ref{eq:omegaG}), would have produced the usual bell-shaped curve
\cite{Pigeau2012}, which corresponds to a diminution of
$\omega(0,H_z)$ when the tip moves away from the nanodot axis. The
behavior for $\lambda=0.5$ is in excellent agreement with the
experimental data, both in the amplitude of the NL frequency shift and
in the position of the compensation point of the force.

We have then repeated the analysis by varying $\delta_z$ in the
experimental error bar range: $\pm 0.2$ around the nominal value. Fit
of the data by a parabola would lead to larger (smaller) values of
$\lambda$ depending if the amplitude of the shift decreases
(increases). This procedure yields an uncertainty window of 30\% for
the determination of $\lambda$, shown as a shaded area in
FIG.\ref{FIG4}a. Our fitting analysis did not account for higher order
corrections in Eq.(\ref{eq:potential}). In FIG.\ref{FIG4}a, the
curvature increases with the displacement distance. Inclusion in the
fit of terms in $ \left| \bm{X} \right|^4$ would have decrease the
value of $\lambda$ by about one standard deviation. As an additional
check, we have performed a simulation of the expected
$\omega(X_0,H_z)$ for our nanodot using a mesh-size of 2.3~nm and a
GPU-accelerated micromagnetic code \cite{Vansteenkiste2011a}. The
result is shown as crosses in FIG.\ref{FIG4}a, demonstrating that our
determination of $\lambda$ is in quantitative agreement with numerical
simulations. It is also in agreement with the result obtained by
Dussaux \textit{et al.}  from micromagnetic simulations performed in
the limit $\xi \gg X_0$ on a thinner dot with approximately the same
radius \cite{Dussaux2012}.

In summary, using an MRFM, we have measured quantitatively the
anharmonicity coefficient $\lambda=+0.5\pm 0.15$ produced by the
depolarisation field of a vortex in a planar nanodot \footnote{A
  recent preprint posted during the review process reports a new
  analytical expression for $\lambda$ compatible with our measurement
  \cite{Metlov2013}}.  From a fundamental perspective, it is
interesting to note that the obtained value (dipole-dominated) is
about twice smaller than the $\lambda=1$ predicted by the local
easy-plane model \cite{Ivanov1998}. Further work is required to check
if the value is independent of the out-of-plane external magnetic
field $\bm{H}_0$ or the dot aspect ratio, $R/t$, in particular around
the line of the vortex state stability $\kappa(R,t)=0$
\cite{Guslienko2008}. Finally, we mention that foldover experiments
performed on the same nanodot at $\bm X_0=0$ produce a red shift of
the gyrotropic frequency (regime $\bm X_0 \ll \xi$), which is opposite
with respect of the sign of $\lambda > 0$. This finding suggests that
the NL frequency shift observed in the foldover of the resonant curve
is not dominated by the anharmonicity of the magnetostatic potential,
but perhaps by the NL damping \cite{Pigeau2011}.

\begin{acknowledgments}
  This research was partly supported by the EU grant MOSAIC
  (ICT-FP7-317950), the French ANR Grant MARVEL
  (ANR-2010-JCJC-0410-01), the Spanish MEC Grants PIB2010US-00153 and
  FIS2010-20979-C02-01. S.S., K.G. and O.S. acknowledge support from
  the Marie Curie grant AtomicFMR (IEF-301656), from the IKERBASQUE
  and from the UPV/EHU, respectively.

\end{acknowledgments}



\begin{thebibliography}{10}%
\makeatletter
\providecommand \@ifxundefined [1]{%
 \ifx #1\undefined \expandafter \@firstoftwo
 \else \expandafter \@secondoftwo
\fi
}%
\providecommand \@ifnum [1]{%
 \ifnum #1\expandafter \@firstoftwo
 \else \expandafter \@secondoftwo
\fi
}%
\providecommand \enquote [1]{``#1''}%
\providecommand \bibnamefont  [1]{#1}%
\providecommand \bibfnamefont [1]{#1}%
\providecommand \citenamefont [1]{#1}%
\providecommand\href[0]{\@sanitize\@href}%
\providecommand\@href[1]{\endgroup\@@startlink{#1}\endgroup\@@href}%
\providecommand\@@href[1]{#1\@@endlink}%
\providecommand \@sanitize [0]{\begingroup\catcode`\&12\catcode`\#12\relax}%
\@ifxundefined \pdfoutput {\@firstoftwo}{%
 \@ifnum{\z@=\pdfoutput}{\@firstoftwo}{\@secondoftwo}%
}{%
 \providecommand\@@startlink[1]{\leavevmode}%
 \providecommand\@@endlink[0]{}%
}{%
 \providecommand\@@startlink[1]{%
  \leavevmode
  \pdfstartlink
   attr{/Border[0 0 1 ]/H/I/C[0 1 1]}%
   user{/Subtype/Link/A<</Type/Action/S/URI/URI(#1)>>}%
  \relax
 }%
 \providecommand\@@endlink[0]{\pdfendlink}%
}%
\providecommand \url  [0]{\begingroup\@sanitize \@url }%
\providecommand \@url [1]{\endgroup\@href {#1}{\urlprefix}}%
\providecommand \urlprefix [0]{URL }%
\providecommand \Eprint[0]{\href }%
\@ifxundefined \urlstyle {%
  \providecommand \doi [1]{doi:\discretionary{}{}{}#1}%
}{%
  \providecommand \doi [0]{doi:\discretionary{}{}{}\begingroup
  \urlstyle{rm}\Url }%
}%
\providecommand \doibase [0]{http://dx.doi.org/}%
\providecommand \Doi[1]{\href{\doibase#1}}%
\providecommand \bibAnnote [3]{%
  \BibitemShut{#1}%
  \begin{quotation}\noindent
    \textsc{Key:}\ #2\\\textsc{Annotation:}\ #3%
  \end{quotation}%
}%
\providecommand \bibAnnoteFile [2]{%
  \IfFileExists{#2}{\bibAnnote {#1} {#2} {\input{#2}}}{}%
}%
\providecommand \typeout [0]{\immediate \write \m@ne }%
\providecommand \selectlanguage [0]{\@gobble}%
\providecommand \bibinfo [0]{\@secondoftwo}%
\providecommand \bibfield [0]{\@secondoftwo}%
\providecommand \translation [1]{[#1]}%
\providecommand \BibitemOpen[0]{}%
\providecommand \bibitemStop [0]{}%
\providecommand \bibitemNoStop [0]{.\EOS\space}%
\providecommand \EOS [0]{\spacefactor3000\relax}%
\providecommand \BibitemShut [1]{\csname bibitem#1\endcsname}%
\bibitem{Slavin2009}%
  \BibitemOpen
  \bibfield{author}{%
  \bibinfo {author} {\bibfnamefont{A.}~\bibnamefont{Slavin}}\ and\ \bibinfo
  {author} {\bibfnamefont{V.}~\bibnamefont{Tiberkevich}},\ }%
  \bibfield{journal}{%
  \Doi{10.1109/TMAG.2008.2009935}{\bibinfo {journal} {Ieee Transactions On
  Magnetics}}\ }%
  \textbf{\bibinfo {volume} {45}},\ \bibinfo {pages} {1875} (\bibinfo {month}
  {Apr.}\ \bibinfo {year} {2009})%
  \bibAnnoteFile{NoStop}{Slavin2009}%
\bibitem{Mohseni2013}%
  \BibitemOpen
  \bibfield{author}{%
  \bibinfo {author} {\bibfnamefont{S.~M.}\ \bibnamefont{Mohseni}}, \bibinfo
  {author} {\bibfnamefont{S.~R.}\ \bibnamefont{Sani}}, \bibinfo {author}
  {\bibfnamefont{J.}~\bibnamefont{Persson \textit{et al.}}}, \bibinfo
  {author}, \ }%
  \bibfield{journal}{%
  \bibinfo {journal} {Science}\ }%
  \textbf{\bibinfo {volume} {339}},\ \bibinfo {pages} {1295} (\bibinfo {month}
  {Mar.}\ \bibinfo {year} {2013})%
  \bibAnnoteFile{NoStop}{Mohseni2013}%
\bibitem{Pribiag2007}%
  \BibitemOpen
  \bibfield{author}{%
  \bibinfo {author} {\bibfnamefont{V.~S.}\ \bibnamefont{{Pribiag}}}, \bibinfo
  {author} {\bibfnamefont{I.~N.}\ \bibnamefont{{Krivorotov}}}, \bibinfo
  {author} {\bibfnamefont{G.~D.}\ \bibnamefont{{Fuchs}}}, \bibinfo {author}
  {\bibfnamefont{P.~M.}\ \bibnamefont{{Braganca}}}, \bibinfo {author}
  {\bibfnamefont{O.}~\bibnamefont{{Ozatay}}}, \bibinfo {author}
  {\bibfnamefont{J.~C.}\ \bibnamefont{{Sankey}}}, \bibinfo {author}
  {\bibfnamefont{D.~C.}\ \bibnamefont{{Ralph}}},\ and\ \bibinfo {author}
  {\bibfnamefont{R.~A.}\ \bibnamefont{{Buhrman}}},\ }%
  \bibfield{journal}{%
  \Doi{10.1038/nphys619}{\bibinfo {journal} {Nature Physics}}\ }%
  \textbf{\bibinfo {volume} {3}},\ \bibinfo {pages} {498} (\bibinfo {month}
  {Jul.}\ \bibinfo {year} {2007}),\
  \bibAnnoteFile{NoStop}{Pribiag2007}%
\bibitem{Locatelli2011}%
  \BibitemOpen
  \bibfield{author}{%
  \bibinfo {author} {\bibfnamefont{N.}~\bibnamefont{Locatelli}}, \bibinfo
  {author} {\bibfnamefont{V.~V.}\ \bibnamefont{Naletov}}, \bibinfo {author}
  {\bibfnamefont{J.}~\bibnamefont{Grollier}}, \bibinfo {author}
  {\bibfnamefont{G.}~\bibnamefont{de~Loubens}}, \bibinfo {author}
  {\bibfnamefont{V.}~\bibnamefont{Cros}}, \bibinfo {author}
  {\bibfnamefont{C.}~\bibnamefont{Deranlot}}, \bibinfo {author}
  {\bibfnamefont{C.}~\bibnamefont{Ulysse}}, \bibinfo {author}
  {\bibfnamefont{G.}~\bibnamefont{Faini}}, \bibinfo {author}
  {\bibfnamefont{O.}~\bibnamefont{Klein}},\ and\ \bibinfo {author}
  {\bibfnamefont{A.}~\bibnamefont{Fert}},\ }%
  \bibfield{journal}{%
  \Doi{10.1063/1.3553771}{\bibinfo {journal} {Applied Physics Letters}}\ }%
  \textbf{\bibinfo {volume} {98}},\ \bibinfo {pages} {062501} (\bibinfo {month}
  {Feb.}\ \bibinfo {year} {2011})%
  \bibAnnoteFile{NoStop}{Locatelli2011}%
\bibitem{Pigeau2010}%
  \BibitemOpen
  \bibfield{author}{%
  \bibinfo {author} {\bibfnamefont{B.}~\bibnamefont{Pigeau}}, \bibinfo {author}
  {\bibfnamefont{G.}~\bibnamefont{de~Loubens}}, \bibinfo {author}
  {\bibfnamefont{O.}~\bibnamefont{Klein}}, \bibinfo {author}
  {\bibfnamefont{A.}~\bibnamefont{Riegler}}, \bibinfo {author}
  {\bibfnamefont{F.}~\bibnamefont{Lochner}}, \bibinfo {author}
  {\bibfnamefont{G.}~\bibnamefont{Schmidt}}, \bibinfo {author}
  {\bibfnamefont{L.~W.}\ \bibnamefont{Molenkamp}}, \bibinfo {author}
  {\bibfnamefont{V.~S.}\ \bibnamefont{Tiberkevich}},\ and\ \bibinfo {author}
  {\bibfnamefont{A.~N.}\ \bibnamefont{Slavin}},\ }%
  \bibfield{journal}{%
  \Doi{10.1063/1.3373833}{\bibinfo {journal} {Applied Physics Letters}}\ }%
  \textbf{\bibinfo {volume} {96}},\ \bibinfo {pages} {132506} (\bibinfo {month}
  {Mar.}\ \bibinfo {year} {2010})%
  \bibAnnoteFile{NoStop}{Pigeau2010}%
\bibitem{Dussaux2012}%
  \BibitemOpen
  \bibfield{author}{%
  \bibinfo {author} {\bibfnamefont{A.}~\bibnamefont{Dussaux}}, \bibinfo
  {author} {\bibfnamefont{A.~V.}\ \bibnamefont{Khvalkovskiy}}, \bibinfo
  {author} {\bibfnamefont{P.}~\bibnamefont{Bortolotti}}, \bibinfo {author}
  {\bibfnamefont{J.}~\bibnamefont{Grollier}}, \bibinfo {author}
  {\bibfnamefont{V.}~\bibnamefont{Cros}},\ and\ \bibinfo {author}
  {\bibfnamefont{A.}~\bibnamefont{Fert}},\ }%
  \bibfield{journal}{%
  \Doi{10.1103/PhysRevB.86.014402}{\bibinfo {journal} {Physical Review B}}\ }%
  \textbf{\bibinfo {volume} {86}},\ \bibinfo {pages} {014402} (\bibinfo {month}
  {Jul.}\ \bibinfo {year} {2012})%
  \bibAnnoteFile{NoStop}{Dussaux2012}%
\bibitem{Novosad2005}%
  \BibitemOpen
  \bibfield{author}{%
  \bibinfo {author} {\bibfnamefont{V.}~\bibnamefont{Novosad}}, \bibinfo
  {author} {\bibfnamefont{F.~Y.}\ \bibnamefont{Fradin}}, \bibinfo {author}
  {\bibfnamefont{P.~E.}\ \bibnamefont{Roy}}, \bibinfo {author}
  {\bibfnamefont{K.~S.}\ \bibnamefont{Buchanan}}, \bibinfo {author}
  {\bibfnamefont{K.~Y.}\ \bibnamefont{Guslienko}},\ and\ \bibinfo {author}
  {\bibfnamefont{S.~D.}\ \bibnamefont{Bader}},\ }%
  \bibfield{journal}{%
  \Doi{10.1103/PhysRevB.72.024455}{\bibinfo {journal} {Phys. Rev. B}}\ }%
  \textbf{\bibinfo {volume} {72}},\ \bibinfo {pages} {024455} (\bibinfo {month}
  {Jul}\ \bibinfo {year} {2005}),\
  \bibAnnoteFile{NoStop}{Novosad2005}%
\bibitem{Drews2012}%
  \BibitemOpen
  \bibfield{author}{%
  \bibinfo {author} {\bibfnamefont{A.}~\bibnamefont{Drews}}, \bibinfo {author}
  {\bibfnamefont{B.}~\bibnamefont{Kr\"uger}}, \bibinfo {author}
  {\bibfnamefont{G.}~\bibnamefont{Selke}}, \bibinfo {author}
  {\bibfnamefont{T.}~\bibnamefont{Kamionka}}, \bibinfo {author}
  {\bibfnamefont{A.}~\bibnamefont{Vogel}}, \bibinfo {author}
  {\bibfnamefont{M.}~\bibnamefont{Martens}}, \bibinfo {author}
  {\bibfnamefont{U.}~\bibnamefont{Merkt}}, \bibinfo {author}
  {\bibfnamefont{D.}~\bibnamefont{M\"oller}},\ and\ \bibinfo {author}
  {\bibfnamefont{G.}~\bibnamefont{Meier}},\ }%
  \bibfield{journal}{%
  \Doi{10.1103/PhysRevB.85.144417}{\bibinfo {journal} {Phys. Rev. B}}\ }%
  \textbf{\bibinfo {volume} {85}},\ \bibinfo {pages} {144417} (\bibinfo {month}
  {Apr}\ \bibinfo {year} {2012}),\
  \bibAnnoteFile{NoStop}{Drews2012}%
\bibitem{Pigeau2011}%
  \BibitemOpen
  \bibfield{author}{%
  \bibinfo {author} {\bibfnamefont{B.}~\bibnamefont{Pigeau}}, \bibinfo {author}
  {\bibfnamefont{G.}~\bibnamefont{de~Loubens}}, \bibinfo {author}
  {\bibfnamefont{O.}~\bibnamefont{Klein}}, \bibinfo {author}
  {\bibfnamefont{A.}~\bibnamefont{Riegler}}, \bibinfo {author}
  {\bibfnamefont{F.}~\bibnamefont{Lochner}}, \bibinfo {author}
  {\bibfnamefont{G.}~\bibnamefont{Schmidt}},\ and\ \bibinfo {author}
  {\bibfnamefont{L.~W.}\ \bibnamefont{Molenkamp}},\ }%
  \bibfield{journal}{%
  \Doi{10.1038/NPHYS1810}{\bibinfo {journal} {Nature Physics}}\ }%
  \textbf{\bibinfo {volume} {7}},\ \bibinfo {pages} {26} (\bibinfo {month}
  {Jan.}\ \bibinfo {year} {2011})%
  \bibAnnoteFile{NoStop}{Pigeau2011}%
\bibitem{Gaididei2010}%
  \BibitemOpen
  \bibfield{author}{%
  \bibinfo {author} {\bibfnamefont{Y.}~\bibnamefont{Gaididei}}, \bibinfo
  {author} {\bibfnamefont{V.~P.}\ \bibnamefont{Kravchuk}},\ and\ \bibinfo
  {author} {\bibfnamefont{D.~D.}\ \bibnamefont{Sheka}},\ }%
  \bibfield{journal}{%
  \Doi{10.1002/qua.22253}{\bibinfo {journal} {International Journal of Quantum
  Chemistry}}\ }%
  \textbf{\bibinfo {volume} {110}},\ \bibinfo {pages} {83} (\bibinfo {month}
  {Jan.}\ \bibinfo {year} {2010})%
  \bibAnnoteFile{NoStop}{Gaididei2010}%
\bibitem{Chen2012}%
  \BibitemOpen
  \bibfield{author}{%
  \bibinfo {author} {\bibfnamefont{T.~Y.}\ \bibnamefont{Chen}}, \bibinfo
  {author} {\bibfnamefont{M.~J.}\ \bibnamefont{Erickson}}, \bibinfo {author}
  {\bibfnamefont{P.~A.}\ \bibnamefont{Crowell}},\ and\ \bibinfo {author}
  {\bibfnamefont{C.}~\bibnamefont{Leighton}},\ }%
  \bibfield{journal}{%
  \Doi{10.1103/PhysRevLett.109.097202}{\bibinfo {journal} {Phys. Rev. Lett.}}\
  }%
  \textbf{\bibinfo {volume} {109}},\ \bibinfo {pages} {097202} (\bibinfo
  {month} {Aug}\ \bibinfo {year} {2012}),\
  \bibAnnoteFile{NoStop}{Chen2012}%
\bibitem{Burgess2013}%
  \BibitemOpen
  \bibfield{author}{%
  \bibinfo {author} {\bibfnamefont{J.~A.~J.}\ \bibnamefont{Burgess}}, \bibinfo
  {author} {\bibfnamefont{A.~E.}\ \bibnamefont{Fraser}}, \bibinfo {author}
  {\bibfnamefont{F.~F.}\ \bibnamefont{Sani}}, \bibinfo {author}
  {\bibfnamefont{D.}~\bibnamefont{Vick}}, \bibinfo {author}
  {\bibfnamefont{B.~D.}\ \bibnamefont{Hauer}}, \bibinfo {author}
  {\bibfnamefont{J.~P.}\ \bibnamefont{Davis}},\ and\ \bibinfo {author}
  {\bibfnamefont{M.~R.}\ \bibnamefont{Freeman}},\ }%
  \bibfield{journal}{%
  \Doi{10.1126/science.1231390}{\bibinfo {journal} {Science}}\ }%
  \textbf{\bibinfo {volume} {339}},\ \bibinfo {pages} {1051} (\bibinfo {year}
  {2013}),\
  \bibAnnoteFile{NoStop}{Burgess2013}%
\bibitem{Klein2008}%
  \BibitemOpen
  \bibfield{author}{%
  \bibinfo {author} {\bibfnamefont{O.}~\bibnamefont{Klein}}, \bibinfo {author}
  {\bibfnamefont{G.}~\bibnamefont{de~Loubens}}, \bibinfo {author}
  {\bibfnamefont{V.~V.}\ \bibnamefont{Naletov}}, \bibinfo {author}
  {\bibfnamefont{F.}~\bibnamefont{Boust}}, \bibinfo {author}
  {\bibfnamefont{T.}~\bibnamefont{Guillet}}, \bibinfo {author}
  {\bibfnamefont{H.}~\bibnamefont{Hurdequint}}, \bibinfo {author}
  {\bibfnamefont{A.}~\bibnamefont{Leksikov}}, \bibinfo {author}
  {\bibfnamefont{A.~N.}\ \bibnamefont{Slavin}}, \bibinfo {author}
  {\bibfnamefont{V.~S.}\ \bibnamefont{Tiberkevich}},\ and\ \bibinfo {author}
  {\bibfnamefont{N.}~\bibnamefont{Vukadinovic}},\ }%
  \bibfield{journal}{%
  \Doi{10.1103/PhysRevB.78.144410}{\bibinfo {journal} {Physical Review B}}\ }%
  \textbf{\bibinfo {volume} {78}},\ \bibinfo {pages} {144410} (\bibinfo {month}
  {Oct.}\ \bibinfo {year} {2008})%
  \bibAnnoteFile{NoStop}{Klein2008}%
\bibitem{Mitsuzuka2012}%
  \BibitemOpen
  \bibfield{author}{%
  \bibinfo {author} {\bibfnamefont{K.}~\bibnamefont{Mitsuzuka}}, \bibinfo
  {author} {\bibfnamefont{D.}~\bibnamefont{Lacour}}, \bibinfo {author}
  {\bibfnamefont{M.}~\bibnamefont{Hehn}}, \bibinfo {author}
  {\bibfnamefont{S.}~\bibnamefont{Andrieu}},\ and\ \bibinfo {author}
  {\bibfnamefont{F.}~\bibnamefont{Montaigne}},\ }%
  \bibfield{journal}{%
  \Doi{10.1063/1.4711219}{\bibinfo {journal} {Applied Physics Letters}}\ }%
  \textbf{\bibinfo {volume} {100}},\ \bibinfo {pages} {192406} (\bibinfo
  {month} {May}\ \bibinfo {year} {2012})%
  \bibAnnoteFile{NoStop}{Mitsuzuka2012}%
\bibitem{Pigeau2012}%
  \BibitemOpen
  \bibfield{author}{%
  \bibinfo {author} {\bibfnamefont{B.}~\bibnamefont{Pigeau}}, \bibinfo {author}
  {\bibfnamefont{C.}~\bibnamefont{Hahn}}, \bibinfo {author}
  {\bibfnamefont{G.}~\bibnamefont{de~Loubens}}, \bibinfo {author}
  {\bibfnamefont{V.~V.}\ \bibnamefont{Naletov}}, \bibinfo {author}
  {\bibfnamefont{O.}~\bibnamefont{Klein}}, \bibinfo {author}
  {\bibfnamefont{K.}~\bibnamefont{Mitsuzuka}}, \bibinfo {author}
  {\bibfnamefont{D.}~\bibnamefont{Lacour}}, \bibinfo {author}
  {\bibfnamefont{M.}~\bibnamefont{Hehn}}, \bibinfo {author}
  {\bibfnamefont{S.}~\bibnamefont{Andrieu}},\ and\ \bibinfo {author}
  {\bibfnamefont{F.}~\bibnamefont{Montaigne}},\ }%
  \bibfield{journal}{%
  \Doi{10.1103/PhysRevLett.109.247602}{\bibinfo {journal} {Phys. Rev. Lett.}}\
  }%
  \textbf{\bibinfo {volume} {109}},\ \bibinfo {pages} {247602} (\bibinfo
  {month} {Dec}\ \bibinfo {year} {2012}),\
  \bibAnnoteFile{NoStop}{Pigeau2012}%
\bibitem{Ivanov2002}%
  \BibitemOpen
  \bibfield{author}{%
  \bibinfo {author} {\bibfnamefont{B.~A.}\ \bibnamefont{Ivanov}}\ and\ \bibinfo
  {author} {\bibfnamefont{G.~M.}\ \bibnamefont{Wysin}},\ }%
  \bibfield{journal}{%
  \Doi{10.1103/PhysRevB.65.134434}{\bibinfo {journal} {Phys. Rev. B}}\ }%
  \textbf{\bibinfo {volume} {65}},\ \bibinfo {pages} {134434} (\bibinfo {month}
  {Mar}\ \bibinfo {year} {2002}),\
  \bibAnnoteFile{NoStop}{Ivanov2002}%
\bibitem{Loubens2009}%
  \BibitemOpen
  \bibfield{author}{%
  \bibinfo {author} {\bibfnamefont{G.}~\bibnamefont{de~Loubens}}, \bibinfo
  {author} {\bibfnamefont{A.}~\bibnamefont{Riegler}}, \bibinfo {author}
  {\bibfnamefont{B.}~\bibnamefont{Pigeau}}, \bibinfo {author}
  {\bibfnamefont{F.}~\bibnamefont{Lochner}}, \bibinfo {author}
  {\bibfnamefont{F.}~\bibnamefont{Boust}}, \bibinfo {author}
  {\bibfnamefont{K.~Y.}\ \bibnamefont{Guslienko}}, \bibinfo {author}
  {\bibfnamefont{H.}~\bibnamefont{Hurdequint}}, \bibinfo {author}
  {\bibfnamefont{L.~W.}\ \bibnamefont{Molenkamp}}, \bibinfo {author}
  {\bibfnamefont{G.}~\bibnamefont{Schmidt}}, \bibinfo {author}
  {\bibfnamefont{A.~N.}\ \bibnamefont{Slavin}}, \bibinfo {author}
  {\bibfnamefont{V.~S.}\ \bibnamefont{Tiberkevich}}, \bibinfo {author}
  {\bibfnamefont{N.}~\bibnamefont{Vukadinovic}},\ and\ \bibinfo {author}
  {\bibfnamefont{O.}~\bibnamefont{Klein}},\ }%
  \bibfield{journal}{%
  \Doi{10.1103/PhysRevLett.102.177602}{\bibinfo {journal} {Physical Review
  Letters}}\ }%
  \textbf{\bibinfo {volume} {102}},\ \bibinfo {pages} {177602} (\bibinfo
  {month} {May}\ \bibinfo {year} {2009})%
  \bibAnnoteFile{NoStop}{Loubens2009}%
\bibitem{Note1}%
  \BibitemOpen
  \bibinfo {note} {At $H_0=6$~kOe, the spins are tilted out-of-plane by about
  20$^\circ $ generating a 7\% decrease of the in-plane component of the
  magnetization outside the vortex core.}%
  \bibAnnoteFile{Stop}{Note1}%
\bibitem{Note2}%
  \BibitemOpen
  \bibinfo {note} {Although the use of piezo-actuators allows ultra-precise
  displacement of the micro-cantilever, the value of $\delta _z$ has inherently
  some uncertainty as it corresponds to the free axis of the cantilever.}%
  \bibAnnoteFile{Stop}{Note2}%
\bibitem{Guslienko2008a}%
  \BibitemOpen
  \bibfield{author}{%
  \bibinfo {author} {\bibfnamefont{K.~Y.}\ \bibnamefont{Guslienko}}, \bibinfo
  {author} {\bibfnamefont{A.~N.}\ \bibnamefont{Slavin}}, \bibinfo {author}
  {\bibfnamefont{V.}~\bibnamefont{Tiberkevich}},\ and\ \bibinfo {author}
  {\bibfnamefont{S.-K.}\ \bibnamefont{Kim}},\ }%
  \bibfield{journal}{%
  \Doi{10.1103/PhysRevLett.101.247203}{\bibinfo {journal} {Phys. Rev. Lett.}}\
  }%
  \textbf{\bibinfo {volume} {101}},\ \bibinfo {pages} {247203} (\bibinfo
  {month} {Dec}\ \bibinfo {year} {2008}),\
  \bibAnnoteFile{NoStop}{Guslienko2008a}%
\bibitem{Guslienko2001}%
  \BibitemOpen
  \bibfield{author}{%
  \bibinfo {author} {\bibfnamefont{K.~Y.}\ \bibnamefont{Guslienko}}, \bibinfo
  {author} {\bibfnamefont{V.}~\bibnamefont{Novosad}}, \bibinfo {author}
  {\bibfnamefont{Y.}~\bibnamefont{Otani}}, \bibinfo {author}
  {\bibfnamefont{H.}~\bibnamefont{Shima}},\ and\ \bibinfo {author}
  {\bibfnamefont{K.}~\bibnamefont{Fukamichi}},\ }%
  \bibfield{journal}{%
  \Doi{10.1103/PhysRevB.65.024414}{\bibinfo {journal} {Phys. Rev. B}}\ }%
  \textbf{\bibinfo {volume} {65}},\ \bibinfo {pages} {024414} (\bibinfo {month}
  {Dec}\ \bibinfo {year} {2001}),\
  \bibAnnoteFile{NoStop}{Guslienko2001}%
\bibitem{Note3}%
  \BibitemOpen
  \bibinfo {note} {All energies can be approximately calculated by running the
  integrand solely on the nanodot volume outside the vortex core region.}%
  \bibAnnoteFile{Stop}{Note3}%
\bibitem{Note4}%
  \BibitemOpen
  \bibinfo {note} {We estimate that $\xi =0.07$ in our experiment, where the
  microwave field strength is $h_\protect \text {rf}=0.6$~Oe.}%
  \bibAnnoteFile{Stop}{Note4}%
\bibitem{Guslienko2002}%
  \BibitemOpen
  \bibfield{author}{%
  \bibinfo {author} {\bibfnamefont{K.~Y.}\ \bibnamefont{Guslienko}}, \bibinfo
  {author} {\bibfnamefont{B.~A.}\ \bibnamefont{Ivanov}}, \bibinfo {author}
  {\bibfnamefont{V.}~\bibnamefont{Novosad}}, \bibinfo {author}
  {\bibfnamefont{Y.}~\bibnamefont{Otani}}, \bibinfo {author}
  {\bibfnamefont{H.}~\bibnamefont{Shima}},\ and\ \bibinfo {author}
  {\bibfnamefont{K.}~\bibnamefont{Fukamichi}},\ }%
  \bibfield{journal}{%
  \Doi{10.1063/1.1450816}{\bibinfo {journal} {Journal of Applied Physics}}\ }%
  \textbf{\bibinfo {volume} {91}},\ \bibinfo {pages} {8037} (\bibinfo {month}
  {May}\ \bibinfo {year} {2002})%
  \bibAnnoteFile{NoStop}{Guslienko2002}%
\bibitem{Guslienko2008}%
  \BibitemOpen
  \bibfield{author}{%
  \bibinfo {author} {\bibfnamefont{K.~Y.}\ \bibnamefont{Guslienko}},\ }%
  \bibfield{journal}{%
  \Doi{10.1166/jnn.2008.003}{\bibinfo {journal} {Journal of Nanoscience and
  Nanotechnology}}\ }%
  \textbf{\bibinfo {volume} {8}},\ \bibinfo {pages} {2745} (\bibinfo {month}
  {Jun.}\ \bibinfo {year} {2008})%
  \bibAnnoteFile{NoStop}{Guslienko2008}%
\bibitem{Guslienko2006}%
  \BibitemOpen
  \bibfield{author}{%
  \bibinfo {author} {\bibfnamefont{K.~Y.}\ \bibnamefont{Guslienko}}, \bibinfo
  {author} {\bibfnamefont{X.~F.}\ \bibnamefont{Han}}, \bibinfo {author}
  {\bibfnamefont{D.~J.}\ \bibnamefont{Keavney}}, \bibinfo {author}
  {\bibfnamefont{R.}~\bibnamefont{Divan}},\ and\ \bibinfo {author}
  {\bibfnamefont{S.~D.}\ \bibnamefont{Bader}},\ }%
  \bibfield{journal}{%
  \Doi{10.1103/PhysRevLett.96.067205}{\bibinfo {journal} {Physical Review
  Letters}}\ }%
  \textbf{\bibinfo {volume} {96}},\ \bibinfo {pages} {067205} (\bibinfo {month}
  {Feb.}\ \bibinfo {year} {2006})%
  \bibAnnoteFile{NoStop}{Guslienko2006}%
\bibitem{Vansteenkiste2011a}%
  \BibitemOpen
  \bibfield{author}{%
  \bibinfo {author} {\bibfnamefont{A.}~\bibnamefont{Vansteenkiste}}\ and\
  \bibinfo {author} {\bibfnamefont{B.}~\bibnamefont{Van~de Wiele}},\ }%
  \bibfield{journal}{%
  \Doi{10.1016/j.jmmm.2011.05.037}{\bibinfo {journal} {Journal of Magnetism and
  Magnetic Materials}}\ }%
  \textbf{\bibinfo {volume} {323}},\ \bibinfo {pages} {2585} (\bibinfo {month}
  {Nov.}\ \bibinfo {year} {2011})%
  \bibAnnoteFile{NoStop}{Vansteenkiste2011a}%
\bibitem{Note5}%
  \BibitemOpen
  \bibinfo {note} {A preprint posted on arXiv during the review process reports
  a new analytical expression for $\lambda $ compatible with our measurement
  \cite {Metlov2013}}%
  \bibAnnoteFile{NoStop}{Note5}%
\bibitem{Ivanov1998}%
  \BibitemOpen
  \bibfield{author}{%
  \bibinfo {author} {\bibfnamefont{B.~A.}\ \bibnamefont{Ivanov}}, \bibinfo
  {author} {\bibfnamefont{H.~J.}\ \bibnamefont{Schnitzer}}, \bibinfo {author}
  {\bibfnamefont{F.~G.}\ \bibnamefont{Mertens}},\ and\ \bibinfo {author}
  {\bibfnamefont{G.~M.}\ \bibnamefont{Wysin}},\ }%
  \bibfield{journal}{%
  \Doi{10.1103/PhysRevB.58.8464}{\bibinfo {journal} {Phys. Rev. B}}\ }%
  \textbf{\bibinfo {volume} {58}},\ \bibinfo {pages} {8464} (\bibinfo {month}
  {Oct}\ \bibinfo {year} {1998}),\
  \bibAnnoteFile{NoStop}{Ivanov1998}%
\bibitem{Metlov2013}%
  \BibitemOpen
  \bibfield{author}{%
  \bibinfo {author} {\bibfnamefont{K.~L.}\ \bibnamefont{{Metlov}}},\ }%
  \bibfield{journal}{%
  \bibinfo {journal} {ArXiv e-prints}}%
   (\bibinfo {month} {Aug.}\ \bibinfo {year} {2013}),\
  \Eprint{http://arxiv.org/abs/1308.0240}{arXiv:1308.0240 [cond-mat.mes-hall]}%
  \bibAnnoteFile{NoStop}{Metlov2013}%
\end{thebibliography}

%

\end{document}